\documentclass[twocolumn,final,aps,prd,longbibliography,nofootinbib,amsfonts,amssymb,amsmath,superscriptaddress]{revtex4-2}

\usepackage[utf8]{inputenc}
\usepackage[T1]{fontenc}
\usepackage{xcolor}
\usepackage{enumerate}
\allowdisplaybreaks[1]
\usepackage{hyperref}
\hypersetup{%
	pdfauthor={Giulia Maniccia and Giovanni Montani and Stefano Antonini},%
	pdftitle={QFT in curved spacetime from quantum gravity: Proper WKB decomposition of the gravitational component}%
}

\newcommand{\ord}[1]{\mathcal{O} \left(#1\right)}
\newcommand{\vgrav}{v_{\mathbf{k}}^{\lambda}}
\newcommand{\uscal}{\phi_{\mathbf{k}}}
\newcommand{\sumgrav}{\sum_{\mathbf{k},\lambda}}
\newcommand{\sumscal}{\sum_{\mathbf{k}}}
\newcommand{\dea}[1]{ \partial_{\alpha} #1 }

\newcommand{\debplus}[1]{\partial_{+} #1}

\newcommand{\debminus}[1]{\partial_- #1}

\newcommand{\degrav}[1]{\partial_{v_{\mathbf{k}}^{\lambda} } #1 }
\newcommand{\degravsq}[1]{\partial_{v_{\mathbf{k}}^{\lambda} }^2 #1 }

\newcommand{\deusq}[1]{\partial_{\phi_{\mathbf{k}} }^2 #1 }
\newcommand{\deT}[1]{\partial_T #1 }

\begin{document}

\title{QFT in Curved Spacetime from Quantum Gravity: proper WKB decomposition of the gravitational component}

\author{Giulia Maniccia}
\email{giulia.maniccia@uniroma1.it}
\affiliation{Physics Department, “La Sapienza” University of Rome, P.le A. Moro 5, 00185 Roma, Italy}
\affiliation{INFN Section of Rome, “La Sapienza” University of Rome, P.le A. Moro 5, 00185 Roma, Italy}

\author{Giovanni Montani}
\email{giovanni.montani@enea.it}
\affiliation{ENEA, FNS Department, C.R. Frascati, Via E. Fermi 45, 00044 Frascati (Roma), Italy} 
\affiliation{Physics Department, “La Sapienza” University of Rome, P.le A. Moro 5, 00185 Roma, Italy}

\author{Stefano Antonini}
\email{santonin@umd.edu}
\affiliation{Maryland Center for Fundamental Physics, University of Maryland, College Park, Maryland 20742, USA}

\date{\today}

\begin{abstract}
    Starting from a reanalysis of previous work, we construct the proper low-energy quantum field theory (QFT) limit of a full quantum gravity theory in the Born-Oppenheimer approach. 
    We separate the gravitational sector into a classical background, given by a vacuum diagonal Bianchi I cosmology, and its quantum perturbations represented by the two graviton degrees of freedom; we further include quantum matter in the form of a test scalar field. We then implement a Born-Oppenheimer separation, where the gravitons and matter play the roles of ``slow'' and ``fast'' quantum components respectively, and perform a WKB expansion in a Planckian parameter. 
    The functional Schr\"odinger evolution for matter is recovered after averaging over quantum-gravitational effects, provided that a condition is imposed on the gravitons' wave functional. Such a condition fixes the graviton dynamics and is equivalent to the purely gravitational Wheeler-DeWitt constraint imposed in previous approaches.
    The main accomplishment of the present work is to clarify that QFT in curved spacetime can be recovered in the low-energy limit of quantum gravity only after averaging over the graviton degrees of freedom, in the spirit of effective field theory. Furthermore, it justifies \emph{a posteriori} the implementation of the gravitational Wheeler-DeWitt equation on the \lq\lq slow'' gravitons' wave functional rather than assuming its validity \emph{a priori}.
\end{abstract}


\maketitle

\section{Introduction}\label{sec:intro}

One of the most striking differences between the gravitational field and other fundamental forces is that, as a consequence of its geometrical nature, the former is an \lq\lq environment'' interaction \cite{bib:misner-gravitation,bib:montani-primordialcosmology,bib:weinberg-gravitation}. 
This peculiarity of the gravitational field is particularly evident when we attempt a canonical quantization of geometrodynamics \cite{bib:dewitt1-1967,*bib:dewitt2-1967,*bib:dewitt3-1967,bib:kuchar-1981,bib:cianfrani-canonicalQuantumGravity,bib:thiemann-book}. In fact, the Hamiltonian vanishes and the quantum evolution appears to be frozen, leading to the so-called \lq\lq problem of time'' in quantum gravity
 \cite{bib:isham-1992,bib:thiemann-book}.
This feature is not altered by the introduction of matter fields, in the presence of which the full gravity-matter Hamiltonian is vanishing. The simple observation that the pure gravity Hamiltonian is no longer zero suggests the possible role of matter as a clock for the gravitational field evolution \cite{bib:kuchar-1991,bib:brown-1995,bib:montani-2002,bib:montanizonetti-2008,bib:montani-cianfrani-zonetti-2009,bib:montani-cianfrani-2009,bib:maniccia-deangelis-2022}.
However, quantum field theory on curved spacetime (QFT-CS) is an established theory \cite{bib:birrel-davies,bib:book-wald-QFT-curved-space,bib:wald-review-1995} which led to a number of intriguing and robust predictions, such as the Unruh effect \cite{bib:crispino-review-unruh-2008} and the Hawking effect \cite{bib:hawking-1975}. 
It is then natural to ask how QFT-CS, which relies on a notion of time, can be recovered from a full timeless quantum gravity theory including matter in the appropriate low-energy limit.

This question was first approached in Ref.~\cite{bib:rubakov-lapchinsky-1979}, where the notion of Tomonaga \lq\lq bubble time'' was introduced. A more robust and physically well-grounded proposal was discussed in Ref.~\cite{bib:vilenkin-1989} using a WKB expansion \cite{bib:landau-quantumMechanics,bib:dunham-1932} in $\hbar$ at zeroth and first order (see also Ref.~\cite{bib:banks-1985}). In Ref.~\cite{bib:vilenkin-1989} the notion of time arose from the matter wave functional's dependence on the quasiclassical gravitational field, which in turn depends at zeroth order on the time coordinate labeling the spacetime foliation (to which we will refer from now on as ``label time'').
The same notion of time was adopted in Ref.~\cite{bib:kiefer-1991} (see also Refs.~\cite{bib:kiefer-2016,bib:kiefer-2018}), where the expansion in a Planckian parameter was considered up to the order where quantum gravity corrections to QFT naturally emerge.
The main merit of Ref.~\cite{bib:kiefer-1991} was to stress how, at such order, a Born-Oppenheimer (BO) \cite{bib:born-1927,bib:bransden} separation between the behavior of the \lq\lq slow'' gravitational variables and the \lq\lq fast'' matter is affected by the serious problem of nonunitarity (see Refs.~\cite{bib:bertoni-1996,bib:venturi-2017,bib:kiefer-2018,bib:kramer-chataignier-2021,bib:digioia-montani-maniccia-2021,bib:maniccia-montani-2022} for possible solutions to this puzzle).

In this paper, we reevaluate the validity of some of the assumptions made in Refs.~\cite{bib:vilenkin-1989,bib:kiefer-1991}; our analysis reformulates on more solid physical grounds the problem of recovering QFT-CS at low energies using a WKB approach.
We consider a mini-superspace model---a Bianchi I vacuum cosmology---with a quantum free scalar field. One major difference (with the analyses \cite{bib:vilenkin-1989,bib:kiefer-1991}) is that we identify a \lq\lq slow'' quantum component in the gravitational sector, represented by independent graviton degrees of freedom. 
Different from Ref.~\cite{bib:vilenkin-1989}, we do not impose \emph{a priori} a separate Wheeler-DeWitt (WDW) equation for the gravitational component only, but rather justify it by using the typical gauge invariance of the BO formulation \cite{bib:kiefer-1991} to have QFT-CS hold in the appropriate limit. The result of our analysis is that the matter dynamics is obtained after averaging over the graviton degrees of freedom, as one would expect in the context of an effective field theory on a quasiclassical background.

This paper is structured as follows. In Sec.~\ref{sec:motivations} we motivate the need for a reformulation of the semiclassical approach to quantum-gravitational corrections by outlining four basic conceptual points. In Sec.~\ref{sec:model} we introduce the model of our interest, whose dynamics is studied in the perturbative WKB scheme in Sec.~\ref{sec:WKBresults}, and also compared with the gravitational WDW equation. Concluding remarks follow in Sec.~\ref{sec:conclusion}.

\section{Motivations for a new scheme}\label{sec:motivations}

We now motivate our analysis by reevaluating some aspects of the proposal developed in Ref.~\cite{bib:vilenkin-1989} (see also Refs.~\cite{bib:kiefer-1991,bib:kiefer-1993,bib:kiefer-1994-review}).
First, we observe that in Ref.~\cite{bib:vilenkin-1989} the separation between a quasiclassical background system and a \lq\lq small'' quantum one was pursued without taking into account the physical nature of the variables. Here, in analogy with Refs.~\cite{bib:rubakov-lapchinsky-1979,bib:kiefer-1991}, we consider a quasiclassical spacetime described by variables $h_A$ (with $A=1,...,n$) and a matter sector described by variables $q_r$ (with $r=1,...,m$).
In both the present discussion and the concrete application below, we focus on a minisuperspace model.
The gravitational component of the super-Hamiltonian in the Arnowitt-Deser-Misner (ADM) formalism \cite{bib:arnowitt-1962} takes the form \cite{bib:vilenkin-1989}
\begin{equation}
	H^g \equiv G_{AB}\, p^A p^B + V(h_A) = 0 \, , 	\label{eq:superHgravity}
\end{equation}
where $p^A$ are the momenta conjugate to $h_A$. Both the minisupermetric $G^{AB}$ and the potential term $V$ are functions of $h_A$, the latter due to the nonvanishing spatial curvature. 
The quantum matter component of the super-Hamiltonian $H^Q$ depends on the matter degrees of freedom $q_r$ as well as on the gravitational variables $h_A$. 

In Ref.~\cite{bib:vilenkin-1989} a BO separation of the quasiclassical and quantum wave functionals was implemented, in which the former is the \lq\lq slow'' and the latter the \lq\lq fast'' component of the coupled system, based on the scale separation $\langle H^Q\rangle/\langle H^g\rangle \sim \hbar$, where $\langle \cdot \rangle$ denotes the expectation value on the respective wave functionals.  
The total wave functional of the gravity+matter system is decomposed as
\begin{equation}
	\Psi (h_A,q_r) = A(h_A) \;e^{iS(h_A)/\hbar}\; \chi(h_A,q_r) \, , 
	\label{eq:VilenkinPsi}
\end{equation}
where the amplitude $A$ and the function $S$ are real, and $\chi$ is associated to quantum matter.

Promoting the two super-Hamiltonian terms to canonical operators $\hat{H}^g$ and $\hat{H}^Q$, the system is quantized \textit{\`a la} Dirac by imposing the following constraints: 
\begin{gather}
	\left( \hat{H}^g + \hat{H}^Q\right)\Psi = 0\,,\label{eq:Vilenkin-total-constr}\\
        \hat{H}^g \,A\,e^{iS/\hbar} = 0\,,
	\label{eq:Vilenkin-grav-constr}
\end{gather}
where Eq.~\eqref{eq:Vilenkin-grav-constr} states that the gravitational component independently satisfies its own WDW equation.
Combining via a WKB expansion in $\hbar$, Eqs.~\eqref{eq:Vilenkin-total-constr}-\eqref{eq:Vilenkin-grav-constr} take the form
\begin{eqnarray}
	G_{AB}\frac{\partial S}{\partial h_A}\frac{\partial S}{\partial h_B} + V(h_A) = 0	\, , \label{eq:Vilenkin-HJ}\\
	G_{AB}\frac{\partial \, }{\partial h_A}\left( A^2 \frac{\partial S}{\partial h_B}\right) = 0	\, , \label{eq:Vilenkin-eq-A}\\
	i\hbar\partial_t\chi = N\hat{H}^Q\chi	\, , \label{eq:Vilenkin-schrod}
\end{eqnarray}
where $N$ is the lapse function, i.e. $dt_s = N(t)dt$, where $t_s$ is the synchronous time. The time derivative in Eq.~\eqref{eq:Vilenkin-schrod} is defined as 
\begin{equation}
	\partial_t\chi \equiv 
	2N G_{AB}\frac{\partial S}{\partial h_A}\frac{\partial \chi}{\partial h_B} = \dot{h}_A\frac{\partial}{\partial h_A}\chi	\, , 
	\label{eq:Vilenkin-time}
\end{equation} 
where in the second equality we made use of the Hamilton equation obtained by varying with respect to $p_A$ (here a dot denotes differentiation with respect to label time). Equation \eqref{eq:Vilenkin-HJ} is of order $\hbar^0$ and corresponds to the Hamilton-Jacobi equation for the classical limit of gravity. 
Both Eqs.~\eqref{eq:Vilenkin-eq-A} and \eqref{eq:Vilenkin-schrod} are obtained\footnote{Unlike Ref.~\cite{bib:vilenkin-1989}, here we adopt the \lq\lq natural'' operator ordering (functions of $h_A$ are always on the left of the corresponding derivatives). This choice, also discussed in Ref.~\cite{bib:kiefer-1991}, has no deep physical implications for the conceptual paradigm.} at order $\hbar$; the former arises from the gravitational WDW equation, while the latter yields the desired QFT dynamics for quantum matter, recovered by simply combining an expansion in $\hbar$ with the BO separation.

Now we are ready to outline four ambiguous points of the approach \cite{bib:vilenkin-1989} which are the main motivations for the present study.
\begin{enumerate}[i)]
    \item \label{point:separate-classical-quantum-dof} The variables $h_A$ do not represent a set of classical gravitational degrees of freedom, because a quantum amplitude $A(h_A)$ is retained at order $\hbar$. Qualitatively, we could write $h_A = h^0_A(t) + \delta h_A$, where $h^0_A(t)$ account for the classical gravitational degrees of freedom (with the dependence on the label time $t$ determined by the Hamilton's equations), while $\delta h_A$ represent quantum corrections of order $\hbar$ to some suitable power.
    Thus, the time differentiation \eqref{eq:Vilenkin-time} should be defined by employing derivatives with respect to $h_A^0$ only, rather than the full quantum variable $h_A$.

    \item \label{point:indep-dyn-gravitons} This also implies that $\delta h_A$ are independent degrees of freedom with respect to $h^0_A(t)$. Therefore, a description of their dynamics is necessary. This is readily understood if we remember that the small metric perturbations of an isotropic universe (whose only degree of freedom is given by the cosmic scale factor $a$) have two scalar, two vector, and two tensor components, at both a classical and a quantum level. These degrees of freedom are independent  from $a$ and are different in number and morphology from the small quantum fluctuations $\delta a$.

    \item \label{point:gauge-BO-gravWDW} Equations \eqref{eq:Vilenkin-eq-A} and \eqref{eq:Vilenkin-schrod} both live at the same order in $\hbar$ and their separation relies on the assumption that it is \textit{a priori} possible to impose the gravitational WDW constraint independently. However, this assumption does not have a physical motivation in the analysis of Ref.~\cite{bib:vilenkin-1989}, and is inconsistent with a pure BO approximation, because it violates its typical gauge invariance. In fact, the BO method separates the whole system into a slow and a fast component, with the wave functional \eqref{eq:VilenkinPsi}. Thus, if we multiply the quantum matter wave functional $\chi$ by a phase depending on $h_A$, the state is invariant provided that we multiply the gravitational component by an inverse phase. This gauge symmetry is broken if we separately impose the gravitational constraint, so that such a procedure appears rather ambiguous. 
    
    \item \label{point:schrod-not-qft} The functional Schr\"odinger equation \eqref{eq:Vilenkin-schrod} is not the right one for quantum matter on a classical curved spacetime, since the matter wave functional $\chi$ depends on the quantum fluctuations of the background $\delta h_A$. This dependence, which was implicitly neglected in Ref.~\cite{bib:vilenkin-1989}, is problematic for the purpose of recovering QFT-CS.   
\end{enumerate}

We would like to remark that the difficulties \ref{point:separate-classical-quantum-dof}), \ref{point:indep-dyn-gravitons}) and \ref{point:schrod-not-qft}) were also present also in Ref.~\cite{bib:kiefer-1991}, while \ref{point:gauge-BO-gravWDW}) was not, because the equation for the quantum-gravitational amplitude $A(h_A)$ was obtained via a gauge condition (see Ref.~\cite{bib:digioia-montani-maniccia-2021} for a comparison of the two approaches in Refs.~\cite{bib:vilenkin-1989} and \cite{bib:kiefer-1991}).

With these motivations, we now reformulate the problem in a Bianchi I cosmological background, obtaining the correct QFT-CS limit without imposing the gravitational constraint and after averaging over quantum-gravitational effects.

\section{Minisuperspace model}\label{sec:model}

Starting from point \ref{point:separate-classical-quantum-dof}) of the previous section, we take the classical cosmological background to be a vacuum diagonal Bianchi I model, which is a homogeneous and spatially flat geometry (the simplest case of the Bianchi classification \cite{bib:landau-classicalfields,bib:montani-primordialcosmology}).
The advantage of this choice over a Friedmann-Lema\^itre-Robertson-Walker model (e.g., in Refs.~\cite{bib:langlois-1994,bib:kiefer-2016-desitter,bib:kiefer-2016,bib:giesel-2020,bib:venturi-2021}) is that, being a vacuum geometry, no scalar or vector perturbations are present \cite{bib:misner-gravitation,bib:weinberg-gravitation}.

In the Misner variables $\alpha$, $\beta_+$ and $\beta_-$ \cite{bib:misner-gravitation,bib:misner-1969}, the line element reads
\begin{equation}
    ds^2 = -N^2(t) dt^2 + e^{\alpha} (e^{\beta})_{ij} dx^i dx^j,
\end{equation}
where $\beta \equiv diag\{ \beta_+ + \sqrt{3}\beta_-, \beta_+ -\sqrt{3}\beta_-, -2\beta_-\}$ is a diagonal traceless matrix. 
The Misner variables depend on the label time $t$ only; $\alpha$ corresponds to the logarithmic volume of the universe, while $\beta_+$ and $\beta_-$ represent the spatial anisotropies.
The supermomentum constraint is identically satisfied and the super-Hamiltonian is
\begin{equation}
     H^{I} (\alpha(t),\beta_{\pm}(t)) = \frac{4}{3M} e^{-\frac{3}{2}\alpha} \left( -p_{\alpha}^2 + p_+^2 + p_-^2 \right), \label{eq:def-superHbianchi}
\end{equation}
where $M = c/32\pi G = c m_{pl}^2/4\hbar$ is a Planckian-order parameter with dimensions of mass over length (with $G$ being the Newton constant and $m_{Pl}$ being the reduced Planck mass). 

According to point \ref{point:indep-dyn-gravitons}), we describe the gravitational fluctuations via tensor perturbations only, as guaranteed by the choice of the vacuum Bianchi I model. Thus the \lq\lq slow'' quantum degrees of freedom $\delta h^A$ correspond to gravitons and are independent from the classical background. 
In the Mukhanov-Sasaki (MS) formalism \cite{bib:sasaki-1986,bib:mukhanov-1987,bib:mukhanov-1992}, the tensor perturbations can be described via the gauge-invariant variables $\vgrav$ in Fourier space ($\lambda$ identifies the two polarization states). For the Bianchi I model \cite{bib:pereira-2007}, the corresponding Hamiltonian (where $N=e^{\alpha}$ in the conformal time $\eta$ gauge) is 
\begin{equation}
    N H^{(v^{\lambda})} = \sum_{\mathbf{k},\lambda} \frac{1}{2} \left[ -\partial_{\vgrav}^2 +\omega_k^2 (\eta) (\vgrav)^2 + \mathcal{V}_{\lambda,\bar{\lambda}} \right]. \label{eq:def-H-gravitons}
\end{equation}
Here each mode $\mathbf{k},\lambda$ behaves as a time-dependent harmonic oscillator with $\omega_{k}^2(\eta) = k^2 -z_{\lambda}''/z_{\lambda}$, where $z_{\lambda}(\eta, k_i)$ is a function of the background metric and $' \equiv\partial_{\eta}$. 
The interaction potential $\mathcal{V}_{\lambda,\bar{\lambda}}$ depends on the shear tensor $\sigma_{ij} = \frac{1}{2} (e^{\beta})'_{ij}$ of the background metric and expresses the mixing of the two polarization modes ($\lambda, \bar{\lambda}$) which takes place due to the anisotropies \cite{bib:pereira-2007} even at the classical level \cite{bib:cho-1995}, differently from isotropic settings.
There is no mixing between scalar and tensor perturbations because we are neglecting the backreaction of the scalar field on the metric (see Refs.~\cite{bib:hu-1978,bib:miedema-1993,bib:agullo-olmedo-2020-theory,bib:agullo-olmedo-2020-numeric} for perturbations in a Bianchi I universe coupled to matter). 

We consider a free test scalar field as the ``fast'' quantum matter sector (e.g. the inflaton field), whose Hamiltonian in the MS formalism takes the form 
\begin{equation}
    N H^{(\phi)} = \sumscal \frac{1}{2} \left[ -\deusq{} +\nu^2_k(\eta)\, (\phi_{\mathbf{k}})^2 \right]. \label{eq:def-H-scal}
\end{equation}
Here, each Fourier mode corresponds to a time-dependent harmonic oscillator with $\nu_{k}^2(\eta) = k^2 -(e^{\alpha})''/ e^{\alpha}$.

The WDW equation for the full model is
\begin{equation}
    \hat{H} \Psi = \left( \hat{H}^{I}  +\hat{H}^{(v^{\lambda})} +\hat{H}^{(\phi)} \right) \Psi = 0 \,,\label{eq:sum-all-H}
\end{equation}
and the wave functional $\Psi$ is assumed to be separable in a BO scheme as 
\begin{equation}
    \Psi = \psi_g(\alpha,\beta_\pm,\vgrav) \,\chi_m(\uscal;\alpha,\beta_\pm,\vgrav)\,.\label{eq:separazione-psi}
\end{equation}
This factorization follows from the assumed difference in energy scale between the matter and gravitational sectors; furthermore, $\psi_g$ is independent of the matter variables $\uscal$ because we assume that the fast quantum sector has a negligible backreaction on the gravitational one. Given the separation~\eqref{eq:separazione-psi}, the WDW equation~\eqref{eq:sum-all-H} is invariant under the transformation:
\begin{equation}
      \psi_g \rightarrow \psi_g e^{-\frac{i}{\hbar} \theta}\,, \quad
      \chi_m \rightarrow e^{\frac{i}{\hbar} \theta} \chi_m \,,\label{eq:phases}
 \end{equation}
where the phase $\theta = \theta (\alpha,\beta_\pm,\vgrav)$ depends on the gravitational variables only. 

As in point \ref{point:gauge-BO-gravWDW}), we will not require the gravitational sector to satisfy the gravitational constraint \emph{a priori}. The gravitons' evolution will instead be derived on physical grounds by requiring the correct QFT dynamics to arise in the appropriate limit and exploiting the gauge invariance \eqref{eq:phases}.

\section{WKB Analysis of the dynamics}\label{sec:WKBresults}

We can now apply the WKB perturbative scheme to our model. We use $1/M$ as the expansion parameter, where $M$ is the (large) Planckian parameter in Eq.~\eqref{eq:def-superHbianchi}. 
This allows us to consistently separate the gravity and matter sectors, in analogy with Refs.~\cite{bib:kiefer-1991,bib:bertoni-1996,bib:kiefer-2016,bib:kiefer-2016-desitter,bib:kiefer-2018,bib:digioia-montani-maniccia-2021,bib:maniccia-montani-2022}. We emphasize that the (semiclassical) WKB expansion in the Planck constant $\hbar$ used in Ref.~\cite{bib:vilenkin-1989} is equivalent to the one used here (see Ref.~\cite{bib:digioia-montani-maniccia-2021} for a detailed discussion on this point).

Expanding up to order $M^0$, the wave function \eqref{eq:separazione-psi} takes the form
\begin{equation}
    \Psi = e^{\frac{i}{\hbar}MS_0}\, e^{\frac{i}{\hbar}(S_1 +\ord{M^{-1}})} \, e^{\frac{i}{\hbar}(Q_1+\ord{M^{-1}})} \,, \label{eq:exp-psi}
\end{equation}
where at leading order $S_0 = S_0(\alpha, \beta_\pm)$. The complex functions $S_n = S_n(\vgrav;\alpha,\beta_\pm)$ and $Q_n = Q_n (\uscal;\alpha,\beta_\pm,\vgrav)$ are associated to the tensor and scalar quantum components of the system, respectively, which must also depend on $\alpha, \beta_{\pm}$.\footnote{Here $S_n$ and $Q_n$ are in general complex functions, whereas in Eq.~\eqref{eq:VilenkinPsi} the exponent $S$ is real-valued and an amplitude $A$ is explicitly extracted.}
The WDW equation \eqref{eq:sum-all-H} applied to Eq.~\eqref{eq:exp-psi} can then be perturbatively examined at each order in $1/M$.
At $\ord{M}$ we obtain
\begin{equation}\label{eq:HJ}
     \frac{4}{3}e^{-\frac{3}{2}\alpha}\, M \left( -\left(\dea{S_0}\right)^2 +\left(\debplus{S_0}\right)^2 +\left(\debminus{S_0}\right)^2 \right) =0,
\end{equation}
which is consistent with the classical Bianchi I solution
\begin{equation}\label{eq:soluz-S0}
    S_0 = k_+ \beta_+ + k_- \beta_- +k_{\alpha} \alpha
\end{equation}
with $k_{\alpha}<0$ corresponding to an expanding universe. 

Let us now introduce the time differentiation operator as in Eq.~\eqref{eq:Vilenkin-time} for $N=e^{\alpha}$, but now constructed using only derivatives with respect to the classical variables $\alpha, \beta_{\pm}$ [this way the issue \ref{point:separate-classical-quantum-dof}) introduced in Sec.~\ref{sec:motivations} does not arise]:
\begin{equation}\label{eq:timederivative}
    -i\hbar \deT{} = \frac{8}{3} e^{-\frac{1}{2}\alpha} \Bigl( \dea{S_0}\,\dea{} +\debplus{S_0}\,\debplus{} +\debminus{S_0}\,\debminus{} \Bigr)\,.
\end{equation}
Using Eqs.~\eqref{eq:timederivative} and \eqref{eq:soluz-S0}, at $\ord{M^0}$ we find
\begin{equation}
\begin{split}
    -i\hbar &(\deT{e^{\frac{i}{\hbar}S_1}}) e^{\frac{i}{\hbar}Q_1} -i\hbar(\deT{e^{\frac{i}{\hbar}Q_1}}) e^{\frac{i}{\hbar}S_1} \\
    &+ \frac{1}{2}\sumgrav \left[\omega_k^2 (\vgrav)^2+\mathcal{V}_{\lambda,\bar{\lambda}}  -\degravsq{}\right] e^{\frac{i}{\hbar}(S_1+Q_1)}\\
    & +\frac{1}{2}\sumscal \left[\nu_k^2 (\uscal)^2-\deusq{}\right] e^{\frac{i}{\hbar}(S_1+Q_1)}=0.
    \end{split}
    \label{eq:ourWDW}
\end{equation}
To address the problem of the dependence of the quantum matter wave functional on the graviton variables [see point \ref{point:schrod-not-qft}) in Sec.~\ref{sec:motivations}] and in the spirit of effective field theory, we average over quantum-gravitational effects to recover QFT-CS, i.e. a functional Schr\"odinger equation for the quantum matter sector. 
To this end, we label $\Gamma_g = \exp(iS_1/\hbar)$ (which is $\psi_g$ at order $M^0$ only) and multiply Eq. \eqref{eq:ourWDW} by the conjugate $\Gamma_g^* = \exp(-iS_1^*/\hbar)$, obtaining
\begin{equation}\label{eq:psi*-psi-chi}
\begin{split}
    -i\hbar &\deT{} \left( \Gamma_g^* \Gamma_g \chi \right) + i\hbar (\deT{\Gamma_g^*})\; \Gamma_g \chi\\
    &+ \frac{1}{2}\sumgrav \left[\left(\omega_k^2 \;(\vgrav)^2\Gamma_g^* +\mathcal{V}_{\lambda,\bar{\lambda}}\Gamma_g^* -\degravsq{\Gamma_g^*} \right) \Gamma_g \chi \right.\\
    &\left. +\degrav{} \left(2 (\degrav{\Gamma_g^*}) \Gamma_g \chi -\degrav{} \left(\Gamma_g^* \Gamma_g \chi \right) \right)\right] \\
    &+\frac{1}{2}\sumscal \left[ \nu_k^2\; (\uscal)^2 -\deusq{}\right] \Gamma_g^* \Gamma_g \chi =0\,,
\end{split}
\end{equation}
where $\chi=\exp(iQ_1/\hbar)$ also depends on the $\vgrav$.
We can eliminate such a dependence by integrating over the $\vgrav$, thus considering an \lq\lq average effect'' of the gravitons. In doing so, we assume that the wave functionals satisfy appropriate boundary conditions such that 
\begin{equation}\label{eq:border-term}
    \int \prod_{\mathbf{k},\lambda} d\vgrav \;\sumgrav \degrav{} \left(2 (\degrav{\Gamma_g^*}) \Gamma_g \chi -\degrav{} \left(\Gamma_g^* \Gamma_g \chi \right) \right)
\end{equation}
vanishes.
In order to recover the desired Schr\"odinger dynamics, we now use the gauge freedom \eqref{eq:phases} to impose the following condition on $\Gamma_g$:
\begin{equation}\label{eq:gauge-cond-grav}
\begin{split}
    \Gamma_g &\Bigl[ i\hbar \deT{\Gamma_g^*} +\frac{1}{2}\sumgrav \left(\omega_k^2 (\vgrav)^2 +\mathcal{V}_{\lambda,\bar{\lambda}} -\degravsq{} \right) \Gamma_g^* \Bigr] \\
    &=0 \,.
\end{split}
\end{equation}
This is possible, provided that the equation
\begin{equation}
    \begin{split}
    \frac{1}{2\hbar} &\sumgrav\left[ -i \degravsq{\theta} +\hbar^{-1} (\degrav{\theta})^2  -i(\degrav{\theta})\degrav{(\ln \Gamma_g^*)}\right]\\
    &-\deT{\theta} =\frac{1}{2}\sumgrav \left[\omega_k^2 (\vgrav)^2 +\mathcal{V}_{\lambda,\bar{\lambda}} -\degravsq{}\right]\Gamma_g^*\\
    & -i\hbar \deT{(\ln \Gamma_g^*)} \,
    \end{split}
    \label{eq:gaugecond}
\end{equation}
has a solution.\footnote{It is understood that the boundary condition~\eqref{eq:border-term} is imposed in the specific gauge set by Eq.~\eqref{eq:gauge-cond-grav}. }
Equations \eqref{eq:psi*-psi-chi} and \eqref{eq:gauge-cond-grav} then guarantee that the \lq\lq averaged'' quantum matter wave functional
\begin{equation}\label{def:theta-tilde}
    \widetilde{\Theta} (\uscal; \alpha, \beta_+, \beta_-) = \int \prod_{\mathbf{k},\lambda} d\vgrav\; \Gamma_g^* \;\Gamma_g \;e^{\frac{i}{\hbar}Q_1}
\end{equation}
satisfies the functional Schr\"{o}dinger equation
\begin{equation}
    i\hbar \,\deT{\widetilde{\Theta}} = \frac{1}{2} \sumscal \left[\nu_k^2 (\uscal)^2  -\deusq{}\right] \widetilde{\Theta} = N \hat{H}^{(\phi)} \widetilde{\Theta} \,,
\end{equation}
therefore recovering QFT-CS on average. We remark that \eqref{eq:gauge-cond-grav} fixes the independent dynamics of gravitons, so the issue \ref{point:indep-dyn-gravitons}) is also resolved. 

At this point, it is worth briefly discussing the relationship between our analysis and standard QFT on curved spacetime \cite{bib:birrel-davies,bib:book-wald-QFT-curved-space}. In that approach, at the one-loop order of approximation, the semiclassical background metric is sourced by the expectation values associated with the quantum components:
\begin{equation}
G_{\mu\nu}^{(0)} = \frac{8\pi G}{c^4} \left( \langle T^{(m)}_{\mu \nu}\rangle 
+ \langle t^{(g)}_{\mu \nu}\rangle \right)
\, , \label{eq:einsteins-eq-backreaction} 
\end{equation}
where $G_{\mu \nu}^{(0)}$ is the Einstein tensor, while $T^{(m)}_{\mu\nu}$ and $t^{(g)}_{\mu\nu}$ denote the energy-momentum tensors of the (renormalized) quantum matter and graviton contributions, respectively. The last two are in principle of the same order, although the graviton term is often neglected in QFT applications \cite{bib:birrel-davies}. In our WKB approach, both backreaction terms are $1/M$ suppressed at leading order \cite{bib:kiefer-QG} and the background is therefore a purely classical vacuum solution described by Eq.~\eqref{eq:HJ}, i.e., the Bianchi I spacetime. 

The backreaction of the fast (matter) component on the slow one does arise at the next order in the general BO scheme, in the form of an expectation value of the matter Hamiltonian,\footnote{Note that, since we work under the assumption that gravitons are part of the ``slow'' component in the BO approximation, they do not give any contribution to the average over the fast sector.} and it was considered in Ref.~\cite{bib:bertoni-1996}. For another formulation of the quantum matter backreaction on the Bianchi I cosmology, see Ref.~\cite{bib:lewandowski-2012}.
This contribution can be removed from the equation governing the matter dynamics and included instead in the gauge condition (\ref{eq:gauge-cond-grav}) specifying the gravitons' dynamics by a phase rescaling of both the matter and gravitational wave functions \cite{bib:digioia-montani-maniccia-2021}. In our analysis we neglected such an expectation value in the gauge condition based on the assumed separation of energy scales between gravitons and matter, but its inclusion does not alter the final result, which is the recovery of QFT-CS in the appropriate low-energy limit.

\subsection{Comparison with gravitational WDW equation}\label{ssec:gravWDW}

Let us now analyze the WKB dynamics arising when separately imposing the gravitational WDW constraint (as in Ref.~\cite{bib:vilenkin-1989}). In the conformal time gauge, this equation reads
\begin{equation}
\begin{split}           
&\left(\hat{H}^I+\hat{H}^{(v^{\lambda})}\right)^{\dagger}\psi_g^* = \Biggl[ \frac{4}{3M} e^{-\frac{3}{2}\alpha} \left(-p_{\alpha}^2 + p_+^2 + p_-^2\right)^{\dagger}\\
    &+\frac{1}{2}e^{-\alpha}\sumgrav \left(- \degravsq{} +\omega^2_k (\vgrav)^2+\mathcal{V}_{\lambda,\bar{\lambda}} \right)^{\dagger} \Biggr] \psi_g^*=0\,
\end{split}
\end{equation}
where $\psi_g^* =\exp{(-i(M S_0^* +S_1^*)/\hbar)}$. 
At $\ord{M^0}$ and using the Hamilton-Jacobi solution \eqref{eq:soluz-S0} for $S_0$, which is real-valued, we obtain
\begin{equation}\label{eq:orderm0grav}
\begin{split}
    -\frac{8}{3} &e^{-\frac{3}{2}\alpha} \Bigl( \dea{S_0}\dea{} +\debplus{S_0}\debplus{} +\debminus{S_0}\debminus{}\Bigr)S_1^* \,e^{-\frac{i}{\hbar}S_1^*} \\
    &+\frac{1}{2}e^{-\alpha}\sumgrav \left[ -i\hbar^{-1}\,\degravsq{S_1^*} +\hbar^{-2}  \left(-\degrav{S_1^*}\right)^2\right.\\
    &\left.+ \left(\omega_k^2\; (\vgrav)^2\right)^{\dagger} +\mathcal{V}_{\lambda,\bar{\lambda}}^{\dagger} \right] e^{-\frac{i}{\hbar}S_1^*}=0 .
\end{split}
\end{equation}
From Eq.~\eqref{eq:timederivative}, this reduces to
\begin{equation}\label{eq:wdwgrav}
    -i\hbar \deT{} \Gamma_g^* = \frac{1}{2}\sumgrav \left( -\degravsq{} +\omega_k^2(\vgrav)^2+\mathcal{V}_{\lambda,\bar{\lambda}} \right)^{\dagger} \Gamma_g^*\,.
\end{equation}
The terms on the right-hand side are Hermitian for each mode $\mathbf{k}, \lambda$ separately, so Eq.~\eqref{eq:wdwgrav} multiplied by $\Gamma_g$  coincides with the condition~\eqref{eq:gauge-cond-grav}. 
Thus, the gravitons' dynamics imposed by selecting the gauge~\eqref{eq:gauge-cond-grav} is equivalent to the one following from the gravitational constraint. In other words, requiring on phenomenological grounds that the quantum matter sector follows the Schr\"odinger dynamics implies that the gravitons' wave functional must satisfy Eq.~\eqref{eq:wdwgrav}.

\section{Discussion and conclusions}\label{sec:conclusion}

Our analysis was motivated by some misleading points (presented in Sec.~\ref{sec:motivations}) of the WKB formulation developed in Refs.~\cite{bib:vilenkin-1989,bib:kiefer-1991}. These works investigated how to obtain the standard quantum dynamics of a \lq\lq small'' (or matter) subsystem from the full WDW equation of such degrees of freedom coupled to quasiclassical (or gravitational) ones, in the limit $\hbar\rightarrow 0$ (or $1/M \rightarrow 0$). 
The basic ambiguity of Refs.~\cite{bib:vilenkin-1989,bib:kiefer-1991} is related to the presence of a quantum correction $\delta h_A$ to the classical background degrees of freedom $h^0_A(t)$ (here we considered a homogeneous diagonal Bianchi I cosmology). 
In the original analysis of Ref.~\cite{bib:vilenkin-1989}, the existence of this quantum correction was implicitly assumed, as is clear from the presence of a quantum amplitude $A(h_A)$ computed at first order in $\hbar$; a similar feature was found in the analysis of Ref.~\cite{bib:kiefer-1991} where the expansion parameter was taken to be $1/M$.
 
In order to address the observations and consequential difficulties listed in points \ref{point:separate-classical-quantum-dof})--\ref{point:schrod-not-qft}) of Sec.~\ref{sec:motivations}, we separated \emph{ab initio} the Bianchi I classical background from its first-order quantum perturbations. Since our background is a vacuum geometry, we restricted our analysis to tensorial perturbations, described by graviton variables. 
We demonstrated that the functional Schr\"odinger equation for the matter sector is correctly recovered after averaging over quantum-gravitational effects. 
To obtain this result, predicted by low-energy phenomenology, we had to fix a gauge from Eq.~\eqref{eq:phases} on the gravitons' sector, whose dynamics corresponds to the one dictated by the gravitational WDW equation only.
The possibility to independently impose such constraint was one of the starting assumptions in Ref.~\cite{bib:vilenkin-1989}, although not sufficiently motivated. 
Since the graviton dynamics cannot be regarded as a gauge-dependent feature, the present study justifies \emph{a posteriori} and on physical grounds the assumption that the gravitational constraint simultaneously holds. In Ref.~\cite{bib:vilenkin-1989}, however, such condition would no longer correspond to a gauge choice, simply because the gauge symmetry was broken from the very beginning.

Apart from Refs.~\cite{bib:vilenkin-1989,bib:kiefer-1991}, other related analyses reconstructed a Schr\"odinger dynamics of a subsystem starting from a quantum gravity framework \emph{à la} BO. 
For instance, in Ref.~\cite{bib:rubakov-lapchinsky-1979} a Tomonaga-Schwinger equation for quantum matter was constructed; that approach is similar to the one discussed here, but the gravitational field was treated as purely classical. 
Our separation of the gravitational degrees of freedom into classical and quantum ones could be implemented in the same scheme, with the expected resulting picture being equivalent to our final outcome. 
In Ref.~\cite{bib:kramer-chataignier-2021},
 the nonunitarity issue of the original study \cite{bib:kiefer-2016,bib:kiefer-2016-desitter} on quantum cosmological perturbations was addressed. 
The authors constructed a suitable inner product, in the spirit of a gauge-fixing approach to the definition of the time variable, and recovered a Schr\"odinger dynamics for the scalar and tensor perturbations including quantum gravity corrections. 
Since we limited our attention to the first two expansion orders (where the nonunitarity issue does not arise), there is no direct overlap between this work and the achievements of Ref.~\cite{bib:kramer-chataignier-2021}. However, it would be interesting to combine the ideas of the present paper, in particular the average on the quantum-gravitational degrees of freedom, with the approach used in Ref.~\cite{bib:kramer-chataignier-2021} to calculate higher-order quantum gravity corrections. In this respect, it is worthwhile to clarify that the tensor fluctuations in Refs.~\cite{bib:kiefer-2016,bib:kiefer-2016-desitter,bib:kramer-chataignier-2021} were treated on the same level as the matter degrees of freedom (i.e., as a fast contribution in the BO scheme), whereas in our approach the gravitons are separated in energy scale from matter (i.e., they belong to the slow component). 

The present study should be regarded as a starting point for future developments of the present approach, where the expansion is performed up to the next orders in the WKB parameter.
Constructing the time variable as discussed above, it is natural to expect the nonunitarity problems analyzed in Refs.~\cite{bib:kiefer-1991,bib:kiefer-2018,bib:digioia-montani-maniccia-2021} to still arise at $\ord{M^{-1}}$. However, the situation can be different for distinct choices of the time coordinates; see, e.g., Refs.~\cite{bib:rubakov-lapchinsky-1979,bib:peres-1999,bib:gielen-2021-unitarity,bib:maniccia-montani-2022}. 
In fact, our study clarifies how the BO approximation for the low-energy dynamics of quantum matter is recovered only after adequate separation of the gravitational degrees of freedom into a main classical background plus small quantum fluctuations.

\begin{acknowledgments}
	G. Maniccia thanks the TAsP INFN initiative (Rome 1 section) for support. S.A. is supported by the U.S. Department of Energy, Office of Science, Office of Advanced Scientific Computing Research, Accelerated Research for Quantum Computing program ``FAR-QC''.
\end{acknowledgments}

\bibliography{gravitons}

\end{document}